\documentclass[prl,twocolumn,amsmath,amssymb,superscriptaddress]{revtex4}

\usepackage{hyperref}
\pdfoutput=1
\usepackage{graphicx}
\usepackage{epsfig}
\usepackage{epstopdf}

\usepackage{amsmath}
\usepackage{bm}
\usepackage{amssymb}
\usepackage{textcomp}
\usepackage{ulem}

\usepackage{import}
\usepackage{color}

\usepackage{float}

\begin{document}

\title{Probing electronic excitations in mono- to pentalayer graphene by micro-magneto-Raman spectroscopy}

\author{St\'ephane Berciaud}
\email{stephane.berciaud@ipcms.unistra.fr}
\affiliation{Institut de Physique et Chimie des Mat\'eriaux de Strasbourg and NIE, UMR 7504, Universit\'e de Strasbourg and CNRS, 23 rue du L\oe{}ss, BP43, 67034 Strasbourg Cedex 2, France}

\author{Marek Potemski}
\affiliation{Laboratoire National des Champs Magn\'etiques Intenses, CNRS/UJF/UPS/INSA, Grenoble F-38042, France}

\author{Cl\'ement Faugeras}
\email{clement.faugeras@lncmi.cnrs.fr}
\affiliation{Laboratoire National des Champs Magn\'etiques Intenses, CNRS/UJF/UPS/INSA, Grenoble F-38042, France}


\begin{abstract}
We probe electronic excitations between Landau levels in freestanding $N-$layer graphene over a broad energy range, with unprecedented spectral and spatial resolution, using micro-magneto Raman scattering spectroscopy. A characteristic evolution of electronic bands in up to five Bernal-stacked graphene layers is evidenced and shown to remarkably follow a simple theoretical approach, based on an effective bilayer model. $(N>3)$-layer graphene appear as appealing candidates in the quest for novel phenomena, particularly in the quantum Hall effect regime. Our work paves the way towards minimally-invasive investigations of magneto-excitons in other emerging low-dimensional systems, with a spatial resolution down to 1$~\mu$m.
\end{abstract}

\maketitle

Searching for materials with new functionalities and deepening our knowledge on already identified structures are primary directions in the research on novel two-dimensional systems, which include the family of Bernal-stacked $N-$layer graphene. Each member of this family displays very distinct electronic properties. This is expected from theory~\cite{Latil2006,Partoens2007,Koshino2007,Koshino2011} and confirmed in experiments on graphene and its bilayer, especially when profiting from a conversion of two-dimensional electronic bands into quantum Hall effect  (QHE) systems with characteristically spaced-in-energy  Landau levels~\cite{Novoselov2005b,Zhang2005,Novoselov2006,Sadowski2006,Jiang2007,Henriksen2008}. Of prime interest are high-quality specimens, often fragile and of small lateral size, thus calling for adequate experimental probes. A well-established path towards obtaining intrinsically pure graphene consists in isolating freestanding layers~\cite{Bolotin2008,Du2009,Berciaud2009}. Such high quality comes at the cost of the relatively small area (typically $<100~\mu\rm m^2$) of available samples and their great environmental sensitivity. A series of freestanding $N-$layer graphene ($N-$LG) may be difficult to study with surface probes (e.g. angle resolved photoemission~\cite{Ohta2007,Coletti2014} and scanning tunneling spectroscopies~\cite{Li2009}), as well as with electron transport measurements~\cite{Novoselov2005b,Zhang2005,Novoselov2006}, which require uniform gate tunability, or optical spectroscopy in the far-infrared range, with intrinsically poor spatial resolution~\cite{Sadowski2006,Jiang2007,Henriksen2008,Mak2010,Orlita2010}. Our method of choice is micro-magneto-Raman scattering spectroscopy (MMRSS), which is minimally invasive and combines high sensitivity with a sub-micrometer spatial resolution, characteristic of visible-light techniques. Indeed, Raman scattering spectroscopy has been widely used to study the electronic excitations in solids and particularly in semiconductors, including two-dimensional electron gases in the quantum Hall effect regime~\cite{PinczukQHE}. So far, however, these studies have been limited to probing electronic excitations in the vicinity of the Fermi energy (intraband excitations, in case of QHE systems). In contrast, the electronic Raman scattering response of graphitic materials is much richer. It may involve both intraband and interband (across the neutrality point) contributions, as theoretically predicted~\cite{Kashuba2009,Mucha-Kruczynski2010} and observed in experiments, e.g. on graphite~\cite{Kossacki2011,Ma2014}, graphene on graphite~\cite{Faugeras2011}, and carbon nanotubes~\cite{Farhat2011}. 

Here, employing MMRSS, with transverse magnetic fields $B$ up to 29 T, we were able to directly trace the electronic Raman scattering features associated with symmetric inter Landau levels transitions in freestanding $N-$LG, from mono- to pentalayer graphene. The dispersion of these excitations with the magnetic field is a hallmark of the number of layers and permits a precise determination of electronic dispersions. With increasing N, the band structure of $N-$LG becomes richer and richer, but remains well accounted by very a simple theoretical model.

\begin{figure*}[!htb]
\begin{center}
\includegraphics[scale=0.65]{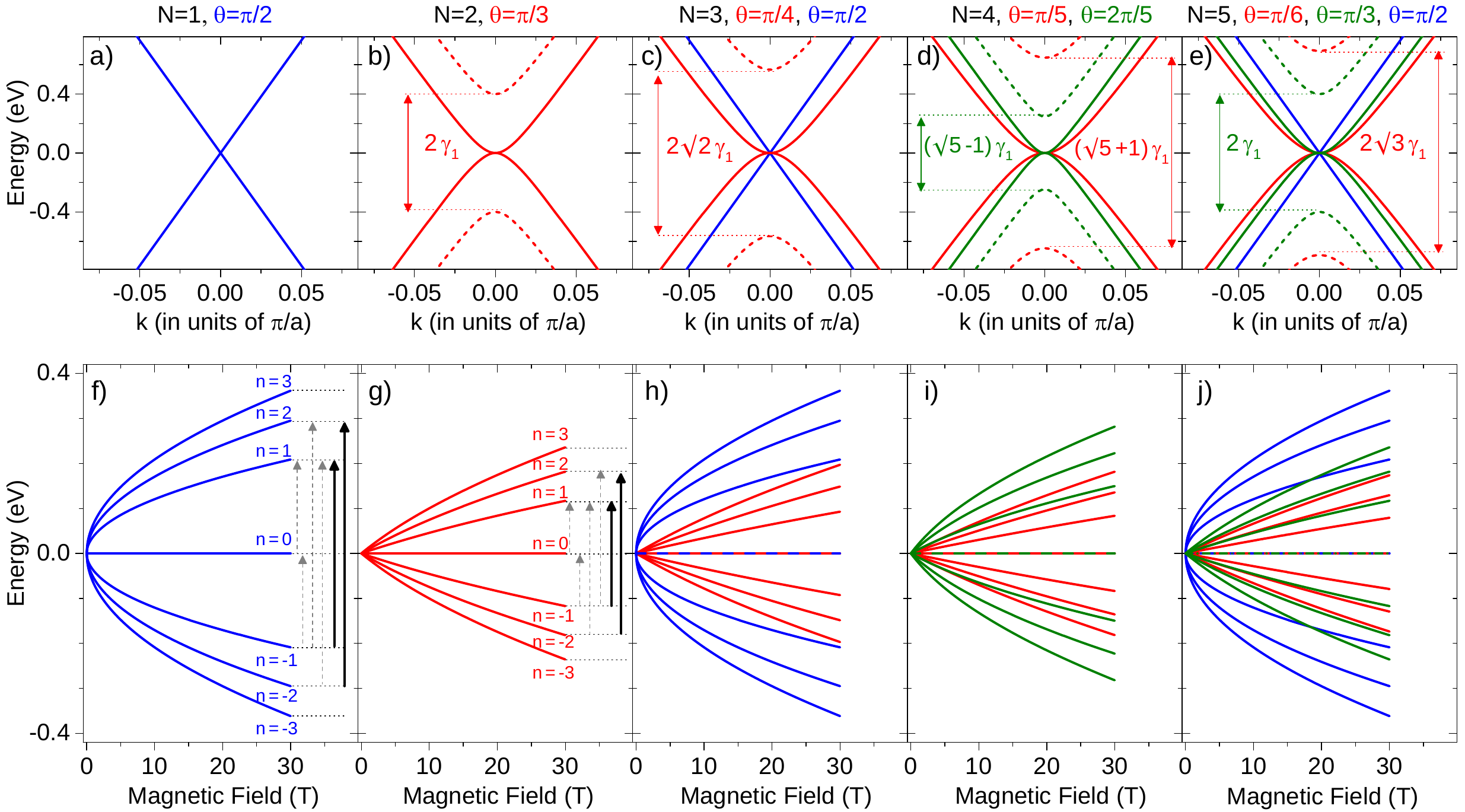}

\caption{Low-energy band structure and Landau levels of $N-$layer graphene. The electronic dispersions of mono-, bi-, tri-, tetra- and pentalayer graphene, obtained from the effective bilayer model described in the text, are shown in a-e, respectively. The corresponding dispersion of the Landau levels arising from the gapless bands are shown in f-j for mono- to pentalayer graphene, respectively. Blue lines in f, h, j, correspond to Landau levels arising for the effective monolayer $\left(\theta=\pi/2\right)$, while red and green lines in g-j correspond to Landau levels arising from effective bilayers obtained at a given quantized value of $\theta\neq  \pi/2$. The first two Raman-active inter-Landau levels electronic transitions $L_{-1,1}^{\theta}$ and $L_{-2,2}^{\theta}$, which satisfy $\delta \left| n \right| =0$ are shown with black arrows in f and g. The first two optical-like inter-Landau levels electronic transitions, $L_{0,1}^{\theta}$ $\left(L_{-1,0}^{\theta}\right)$ and $L_{-2,1}^{\theta}$ $\left(L_{-1,2}^{\theta}\right)$, which satisfy $\delta \left| n \right| =\pm 1$ are shown with dashed grey arrows in f and g. Those excitations are responsible for the magneto-phonon resonances. The calculations are performed with $v_F=1.05\times10^6~\rm m/s$ and $\gamma_1=400~\rm$~meV.}

\label{Fig1}
\end{center}
\end{figure*}

In the latter model, referred to as the \textit{effective bilayer model}~\cite{Latil2006,Koshino2007,Partoens2007,Orlita2009a,Mak2010,Orlita2010}, we start with the electronic bands of graphite, which are very simply simulated using only two tight-binding parameters, namely the nearest neighbor hopping parameter $\gamma_0$, which defines the Fermi velocity $v_F=3/2 a \gamma_0/\hbar$, where $a=0.142~\rm nm$ is the \textsc{c--c} bond length, and the nearest neighbor interlayer coupling constant $\gamma_1$. The band structure of $N-$LG is then derived from two-dimensional cuts in the electronic dispersion of graphite, perpendicular to the transverse momentum $\left(k_z\right)$ direction, at particular values of $\theta=k_z~c/2 =\left(p \pi\right) / \left(N+1\right)$, where $\rm N$ is the number of layers, and $p =\pm 1,\:\pm 2,\dots \pm \left\lfloor (N+1)/2\right\rfloor$, where $\left\lfloor \right\rfloor$ denotes the integer part and $c/2=0.34~\rm nm$ is the interlayer spacing in  bulk graphite. As shown in Figure~\ref{Fig1}a-e, within this model, effective bilayers have formally identical dispersions as that of bilayer graphene, with a rescaled $\tilde{\gamma}_1=2\gamma_1\!\cos\theta$~\cite{Koshino2007,Partoens2007,Orlita2009a,Mak2010}, which corresponds to half the energy gap between their split-off bands. Monolayer graphene, (at $\theta=\pi/2$, \textit{i.e.,} $\tilde{\gamma}_1=0$) displays a well-known linear dispersion, with a slope given by $v_F$.  Bilayer graphene exhibits a set of four hyperbolic bands, obtained at $\theta=\pi/3$, \textit{i.e.}, $\tilde{\gamma}_1=\gamma_1$. Trilayer graphene hosts a pair of graphene-like bands (at $\theta=\pi/2$) and an effective bilayer with $\tilde{\gamma}_1=\sqrt{2}\gamma_1$ at $\theta=\pi/4$. In this framework, tetralayer graphene displays two effective bilayers at $\theta=\pi/5$ and $\theta=2\pi/5$ and pentalayer graphene possesses graphene-like bands at $\theta=\pi/2$ and two sets of effective bilayer bands at $\theta=\pi/6$ and $\theta=\pi/3$.

\begin{figure*}[!ht]
\begin{center}
\includegraphics[scale=0.62]{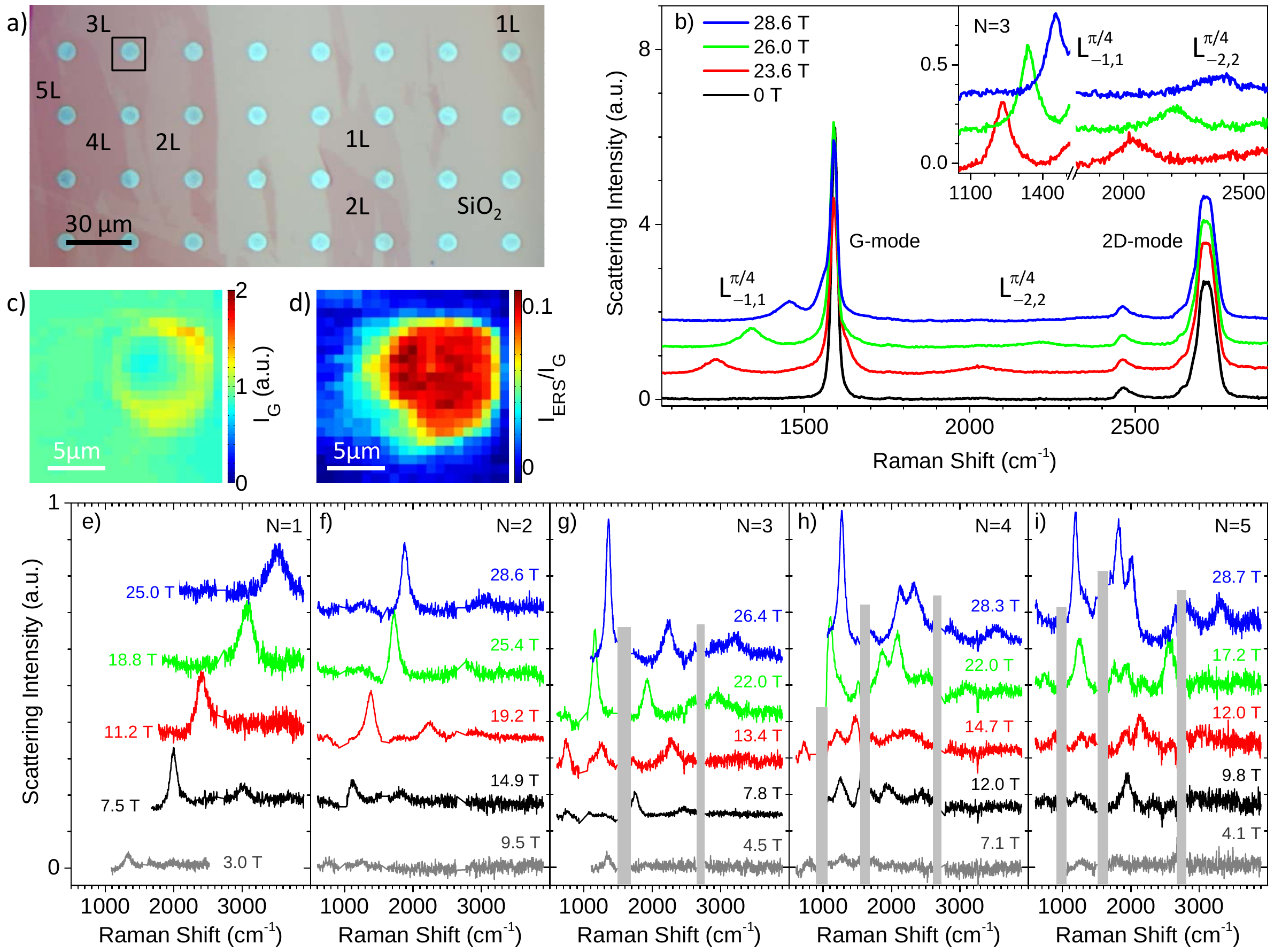}
\caption{Observation of Raman-active electronic excitations in $N-$layer graphene. a) Optical image of a mechanically exfoliated sample showing several
$N-$LG flakes partly freestanding over $8~\rm \mu m$ circular pits spaced by a $30~\rm \mu m$ pitch. b) Raman scattering spectra of a freestanding graphene trilayer at various magnetic fields. The first $\left(L^{\pi/4}_{-1,1}\right)$ and second $\left(L^{\pi/4}_{-2,2}\right)$ Raman active electronic excitations, associated with the effective bilayer, are clearly visible. Their evolution with increasing magnetic field is shown in the inset, which plots the same Raman spectra, after subtraction of the Raman spectrum recorded at $\rm B=0~\rm T$. c, d) Raman maps around the freestanding trilayer graphene region boxed in a), recorded at $\rm B=22~\rm T$. While the integrated intensity of the G-mode feature varies little over the scanned area (c), the ratio of integrated intensity of the $L^{\pi/4}_{-1,1}$  electronic Raman scattering (ERS) feature (around $1150~\rm cm^{-1}$) to that of the G-mode feature shows a high contrast between the freestanding and supported regions (d). e-i) show selected Raman spectra at different magnetic fields, after subtraction of the Raman spectrum recorded at $\rm B=0~\rm T$. The grey vertical bars mask residual contributions from the G- and 2D-mode features in $N-$LG, near $1585~\rm cm^{-1}$, and $2700~\rm cm^{-1}$, respectively, and from the underlying Si substrate, near $1000~\rm cm^{-1}$. Noteworthy, in trilayer graphene (g), the  $L^{\pi/2}_{-1,1}$ and $L^{\pi/2}_{-2,2}$ electronic excitations, associated with the effective monolayer bands, as well as the $L^{\pi/4}_{-3,3}$ feature from the effective bilayer bands, which were not shown in b), are also clearly detectable. The $L^{\pi/2}_{-1,1}$ and $L^{\pi/4}_{-3,3}$ features merge at fields near 29~T (see also Figure~\ref{Fig3}).}

\label{Fig2}
\end{center}
\end{figure*}

In the presence of a transverse magnetic field, Landau quantization occurs and the effective mono- and bilayers give rise to independent sets of Landau levels (LL). The energy  $L^{\theta}_{n}$ $\left(L^{\theta}_{-n}=-L^{\theta}_{n}\right)$ of the $n^{th}$ electron (hole) LL, arising from the gapless bands in an effective bilayer is given by~\cite{McCann2006a,Orlita2010}:

\begin{equation*}
L^{\theta}_{\left|n\right|}=\sqrt{\frac{\tilde{\gamma}_1^2}{2}+(\left|n\right|+\frac{1}{2})E_1^2-\sqrt{\frac{\tilde{\gamma}_1^4}{4}+(\left|n\right|+\frac{1}{2})E_1^2\tilde{\gamma}_1^2+\frac{E_1^4}{4}}},
\end{equation*}

where $E_1=v_F\sqrt{2e\hbar \rm B}$. At $\theta=\pi/2$, one recognizes the well known LL dispersion of monolayer graphene: $L^{\pi/2}_{\left|n\right|}=\sqrt{n}E_1$. For $\theta\neq\pi/2$, a nearly linear scaling of $L^{\theta}_{n}$ with $\rm B$ is expected. The corresponding dispersions are represented in Figure~\ref{Fig1}f-j. The LL structure of $N-$LG becomes increasingly rich as $\rm N$ increases, and crossings between LL arising from distinct subbands appear. As previously demonstrated \cite{Kashuba2009,Mucha-Kruczynski2010,Kossacki2011}, the inter LL transitions which satisfy $\delta \left| n \right| = 0$ (with energy $L^{\theta}_{-n,n}=2\times L^{\theta}_{\left|n\right|}$) are Raman allowed, while "optical-like" transitions, such that $\delta \left| n \right| = \pm 1$ couple to zone center phonons and give rise to the magneto-phonon resonance~\cite{Ando2007a,Goerbig2007,Faugeras2009}.

An optical image of the sample used in our experiments is shown in Figure~\ref{Fig2}a. $N-$layer graphene specimens have been prepared by micro-mechanical exfoliation of natural graphite and deposited onto  SiO$_2$/Si substrates, pre-patterned with an array of $8~\rm \mu m$ diameter circular pits, as in refs.~\cite{Berciaud2009}. The freestanding $N-$layer flakes of interest are readily identified by optical contrast and Raman spectroscopy measurements at $\rm B=0~\rm T$~\cite{Berciaud2009,Berciaud2013,Ferrari2013} (see also supporting information). In this work, only Bernal-stacked $N-$LG were investigated.

The low-temperature micro-magneto-Raman scattering response of suspended $N-$LG has been measured with a custom-made set-up. The excitation laser at $\lambda=514.5 ~\rm nm$ (\textit{i.e.}, a laser photon energy of $2.41~\rm eV$) is injected into a 5 $\mu$m core optical fiber and then focused onto the sample with a high numerical aperture spherical lens. The unpolarized back-scattered light is collected by the same lens, then injected into a 50 $\mu$m optical fiber coupled to a monochromator, equipped with a liquid nitrogen cooled charge coupled device (CCD) array. The excitation laser power was set to $\sim 1~\rm mW$ focused onto a $\sim 1\mu$m diameter spot. The resulting intensity is sufficiently low to avoid significant laser-induced heating and subsequent spectral shifts of the Raman features. The sample was held in a magnetic cryostat at a base temperature of $4~\rm K$ under a residual He pressure, and mounted on X-Y-Z piezo stages, allowing us to move the sample relative to the laser spot with sub-$\mu \rm m$ accuracy. The evolution of the Raman spectrum with magnetic field was then measured on freestanding samples by slowly sweeping the magnetic field from $\rm B=0~\rm T$ to $\rm B=29~\rm T$, while recording spectra on the fly. Thus, each individual spectrum is the integrated response over $\delta  B \sim0.1~\rm T$. We verified, by recording selected spectra at fixed $\rm B$, that this had no impact whatsoever on the linewidth of the electronic Raman features discussed here. Each electronic Raman feature was fit as a function of $\rm B$, using Lorentzian forms.

Let us first consider the illustrative case of trilayer graphene. Figure~\ref{Fig2}b shows selected Raman scattering spectra recorded at $\rm B=0~\rm T$ and at magnetic fields near $25~\rm T$.  Two prominent field-induced changes are observable in these spectra.

\noindent i) Magnetic field-dependent sidebands appear on the low- and high-energy sides of the G-mode feature. At $ B \sim 26~\rm T$, the $L^{\pi/2}_{0,1}$ $\left(L^{\pi/2}_{-1,0}\right)$ and $L^{\pi/4}_{-1,2}$ $\left(L^{\pi/4}_{-2,1}\right)$ have energies close to that of the G-mode phonon ($\sim 1585~\rm cm^{-1}$ or $197~\rm meV$) and give rise to magneto-phonon resonances (MPR)~\cite{Ando2007a,Goerbig2007,Faugeras2009}. Details on MPR in freestanding $N-$LG will be reported elsewhere.

\noindent ii) Dispersive features appear near $1150-1450~\rm cm^{-1}$ and $2000-2400~\rm cm^{-1}$. Their integrated intensity is up to $\sim 10\%$ of that of the G-mode feature and their frequencies increase quasi linearly with $\rm B$, as expected for effective bilayer bands. These features are tentatively assigned to the $L^{\pi/4}_{-1,1}$ and $L^{\pi/4}_{-2,2}$ transitions, respectively.

\begin{figure*}[!ht]
\begin{center}
\includegraphics[scale=0.78]{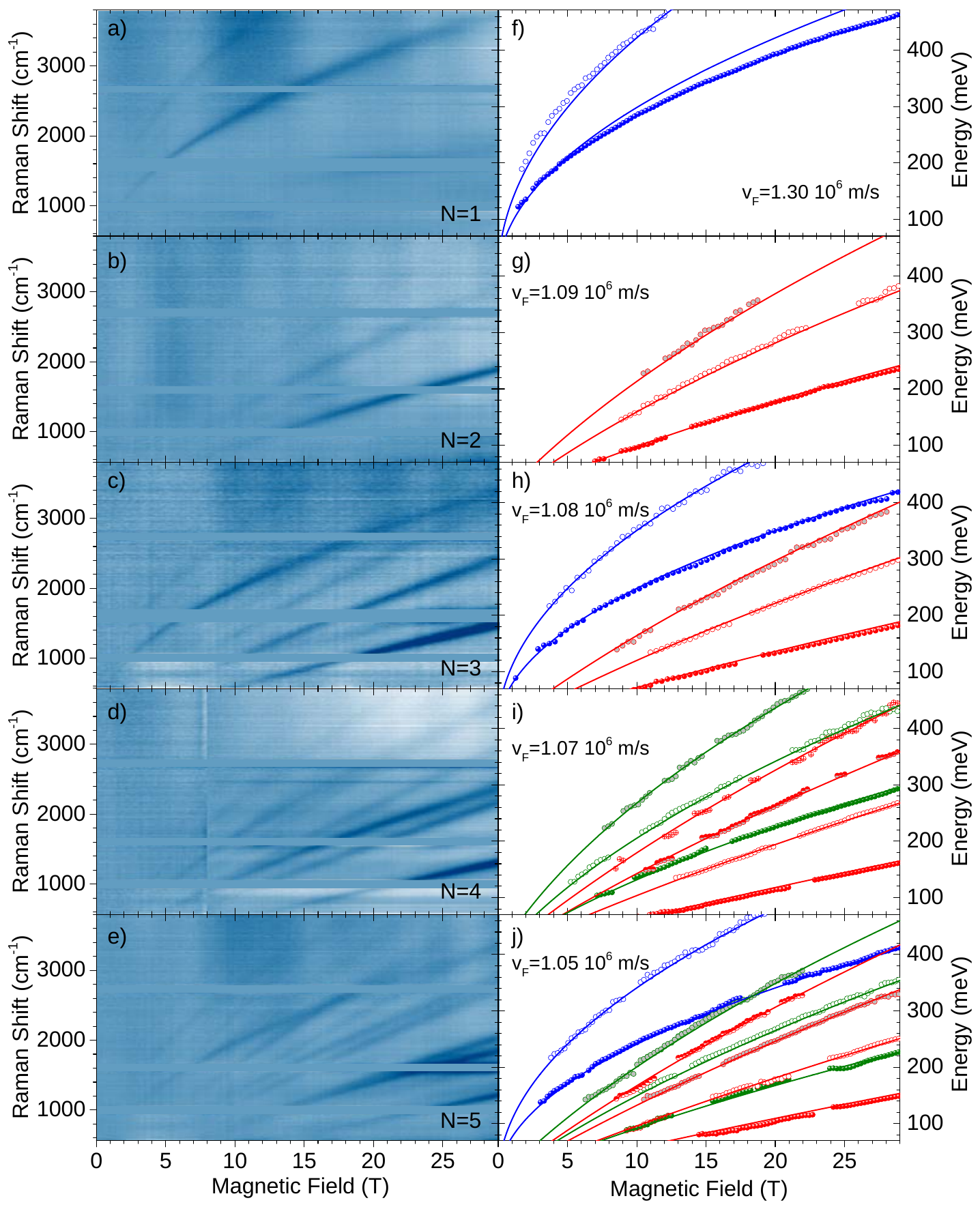}
\caption{Dispersion of the electronic excitations in $N-$layer graphene. a-e) show false-color maps of the micro-magneto-Raman scattering spectra of mono- to pentalayer graphene, as a function of the magnetic field, after subtraction of the spectrum recorded at $\rm B=0~\rm T$. Horizontal light blue bars mask residual contributions from the G- and 2D-mode features in $N-$LG, near $1585~\rm cm^{-1}$, and $2700~\rm cm^{-1}$, respectively, and from the underlying Si substrate, near $1000~\rm cm^{-1}$. The corresponding peak frequencies of the electronic Raman features extracted from Lorentzian fits are shown in f-j). The peak frequencies are determined with an experimental error that is smaller than the symbol size. The solid lines are calculated dispersions of the Landau levels assuming a linear dispersion for monolayer graphene and using the effective bilayer model. The solid lines follow the same color code as in Figure~\ref{Fig1}. A same interlayer hopping parameter of $\gamma_1=400~\rm meV$ was used for all calculations, while the Fermi velocity that best fits our results is indicated in each panel.}

\label{Fig3}
\end{center}
\end{figure*}

 Figure~\ref{Fig2}c depicts a map of the G-mode integrated intensity $I_G$ recorded at $\rm B=22\rm~T$ over the trilayer graphene area boxed in  Figure~\ref{Fig2}a.  A moderate change in $I_G$, can be seen near the edge of the pit, while $I_G$ has similar values at the center of the pit and on the supported region. 
At this field, the $L^{\pi/4}_{-1,1}$ feature appears near $1150~\rm cm^{-1}$. Remarkably, the ratio of integrated intensity of the latter feature to that of the G-mode feature is hardly detectable on supported TLG and reaches up to $\sim 10\%$ on the freestanding area. This demonstrates that: i) the ERS features are strongly damped on supported $N-$LG, presumably due to substrate-induced inhomogeneities and enhanced electron scattering rate, ii) that our experimental approach allows for a spatially resolved and non-contact investigation of electronic excitations in $N-$LG with a spatial resolution down to $\sim 1~\rm \mu m$.

 Figure~\ref{Fig2}e-i displays a set of representative Raman spectra, measured on mono- to pentalayer graphene at different values of the magnetic field. In each $N-$LG specimen, electronic excitations, with different dispersions as a function of the magnetic field, appear. These ERS features are well fit to Lorentzian forms. This readily allows us to extract full widths at half maximum $\Gamma$, which grow moderately as a function of the energy of the electronic excitation. Typical $\Gamma$ are in the range $50~\rm cm^{-1} - 200~\rm cm^{-1}$, \textit{i.e.}, at least three times narrower than those extracted from magneto-absorption measurements on supported monolayer graphene~\cite{Jiang2007}. This corresponds to a quasiparticle lifetime of $\sim 220~\rm fs$ down to $\sim 50~\rm fs$, in line with typical electron scattering times deduced from magneto-transport measurements on high mobility freestanding graphene~\cite{Bolotin2008}. For monolayer graphene, we can easily trace a first ERS feature, which energy increases from $\sim 1300~\rm cm^{-1}$ at $\rm B=3~\rm T$ up to $\sim 3500~\rm cm^{-1}$ at $\rm B=25~\rm T$. This feature is assigned to the $L^{\pi/2}_{-1,1}$ transition. A fainter, second ERS feature, assigned to the $L^{\pi/2}_{-2,2}$ transition emerges from the background near $2000~\rm cm^{-1}$ at $\rm B=3~\rm T$, and its energy increases up to $3600~\rm cm^{-1}$ at $\rm B=11.2~\rm T$. For bilayer graphene, up to three ERS features, assigned to the $L^{\pi/3}_{-1,1}$, $L^{\pi/3}_{-2,2}$ and $L^{\pi/3}_{-3,3}$ are observed. For N$>$2, the magneto-Raman spectra then become increasingly rich.

To clearly visualize electronic excitations, we now consider the contour plot built from the magneto-Raman spectra recorded as a function of $B$ (see Figure~\ref{Fig3}) on a diffraction limited spot. On these maps, the dispersion of the electronic Raman features identified in Figure~\ref{Fig2} appear prominently. The difference between the dispersion of the ERS features in mono- and bilayer graphene is particularly striking.  The high sensitivity of our setup allows us to trace up to 2, 3, 5, 7 and 9 excitations in $N=1$ to $N=5$ layer graphene, respectively. As anticipated theoretically, monolayer-like dispersions are only observed for odd N, while bilayer-like dispersions appear for N$>$1. 

 Figure~\ref{Fig3}e-i, shows the extracted frequency of all the observable ERS features, along with fits, based on the model described above for the corresponding N. Interestingly, the LL dispersions observed in freestanding monolayer graphene deviate significantly from the expected $\sqrt{nB}$ scaling, and suggest average Fermi velocities of $\sim 1.3\times10^6~\rm m/s$, significantly larger than the values of $\sim 1.0-1.1\times10^6~\rm m/s$ from previous magneto-absorption studies on supported graphene~\cite{Orlita2010}. 

First, the relatively high values of $v_F$ observed in monolayer graphene likely stem from the reduction of dielectric screening in freestanding graphene~\cite{Elias2011,Hwang2012}.  Second, the peculiar scaling of $L_{-n,n}^{\pi/2}$ with $n$ and $B$ observed for $N=1$ (see Figure~\ref{Fig3}a,f) is presumably a consequence of many body Coulomb interactions in freestanding graphene. For example, recent electron transport measurements on freestanding graphene devices have revealed a logarithmic divergence of the Fermi velocity in the close vicinity of the Dirac point\cite{Elias2011}. This behavior has been assigned to electron-electron interactions. In our work, from the Raman scattering spectrum measured at $ B=0~ \rm T$, we estimate a residual doping below $\sim 2\times 10^{11}~\rm cm^{-2}$\cite{Berciaud2009,Berciaud2013}, a value at which such interaction effects become significant\cite{Elias2011}.  A comprehensive analysis of the influence of many-body effects on the magneto excitons in freestanding graphene and, subsequently, on the evolution of the Fermi velocity in the presence of a magnetic field~\cite{Shizuya2011}, goes beyond the scope of the present study and will be discussed elsewhere.

In comparison, for $N=2$ to $N=5$, the dispersions of all the observed ERS features resemble the theoretical patterns introduced in Figure~\ref{Fig1}, and are quantitatively reproduced by the effective bilayer model. At a given value of $\theta$, we observe that the $L_{-n,n}^{\theta}$ transition exhibits a steeper dispersion as a function of $B$ for smaller N. Within the effective bilayer framework, and using a fixed value of $\gamma_1=400~\rm meV$, in keeping with previous measurements on bulk graphite~\cite{Orlita2009a,Orlita2010}, this translates into a slight reduction of the Fermi velocity from $v_F=\left( 1.09\pm0.01\right)\times10^6~\rm m/s$ for $N=2$ down to $\left( 1.05\pm0.01\right)\times10^6~\rm m/s$ for $N=5$.  We have also attempted to fit our results using the values of $L_{-n,n}^{\theta}$ derived from the Slonczewski-Weiss-McClure model for the dispersion of bulk graphite~\cite{Nakao1976}. Using several additional fitting parameters, this model provides a marginally better fit to our data, and also suggests a very similar reduction of $v_F$ with increasing N. Our observations also suggest that for $N>1$, the existence of low-energy, gapless electronic bands with finite curvature, and the subsequent finite density of states at the Dirac point at $\rm B=0~\rm T$ and increased degeneracy of the $n=0$ LL at finite $\rm B$, play an essential role in defining the Fermi velocity. Altogether, the apparent decrease of $v_F$ with increasing N is consistent with the value of $v_F=1.02\times10^6~\rm m/s$ observed in bulk graphite~\cite{Orlita2009a} and with recent angle resolved photoemission spectroscopy measurements on graphene monolayers deposited on various substrates~\cite{Hwang2012}.

In conclusion, our results demonstrate the richness of the electronic excitation spectra in $N-$layer graphene and reveals a peculiar dispersion of Landau levels in monolayer graphene, thus opening interesting avenues for the study of many-body effects on its magneto-optical conductivity. The experimental methodology presented here permits the optical probing of electron-hole quasiparticles, from energies as low as $\sim 100~\rm meV$ in monolayer graphene, and down to $\sim 60~\rm meV$ in (N$>$1)-layer graphene, using moderate magnetic fields of a few T. In addition, most of the electronic excitations presented in this work can readily be observed using magnetic fields below $12$~T, accessible with commercially available superconducting magnets. Consequently, table-top micro-magneto Raman scattering experiments, may be developed to unravel electronic excitations in the close vicinity of the charge neutrality point of $N-$layer graphene and to further investigate the electronic dispersion of these systems.
Finally, our work provides an impetus for studies of (N$>$3)-layer graphene in the Quantum Hall effect regime~\cite{Koshino2011}, in which new broken symmetry states~\cite{Taychatanapat2011} can be expected due to extra crossings of Landau level and/or Landau level bunching over a broad range of magnetic field. We also believe that further improvement of the experimental arrangements will allow us to measure the excitations in a close vicinity to the laser line, thus offering an optical probe of small energy gaps, such as, for example, those arising from lifting the spin or valley degeneracy~\cite{Feldman2009}.


\subsection* {Acknowledgments}

We are grateful to M. Orlita, P. Kossacki, and D.M Basko for numerous discussions. We also thank Ivan Breslavetz for technical support and R. Bernard,  S. Siegwald, and H. Majjad for help with sample preparation in the StNano clean room facility. S.B. acknowledges support from the Agence nationale de la recherche (under grant QuanDoGra 12 JS10-001-01), from the CNRS and Universit\'e de Strasbourg and from the LNCMI-CNRS, member of the European Magnetic Field Laboratory (EMFL). M.P. and C.F.  acknowledge support from the ERC-2012-AdG-320590 MOMB project and the EC graphene flagship. 








\onecolumngrid
\newpage
\begin{center}
{\large\textbf{Supporting Information}}
\end{center}

\setcounter{equation}{0}
\setcounter{figure}{0}
\renewcommand{\theequation}{S\arabic{equation}}
\renewcommand{\thefigure}{S\arabic{figure}}

This documents contains five supporting figures with detailed captions.

\begin{figure}[!ht]
\begin{center}
\includegraphics[scale=0.5]{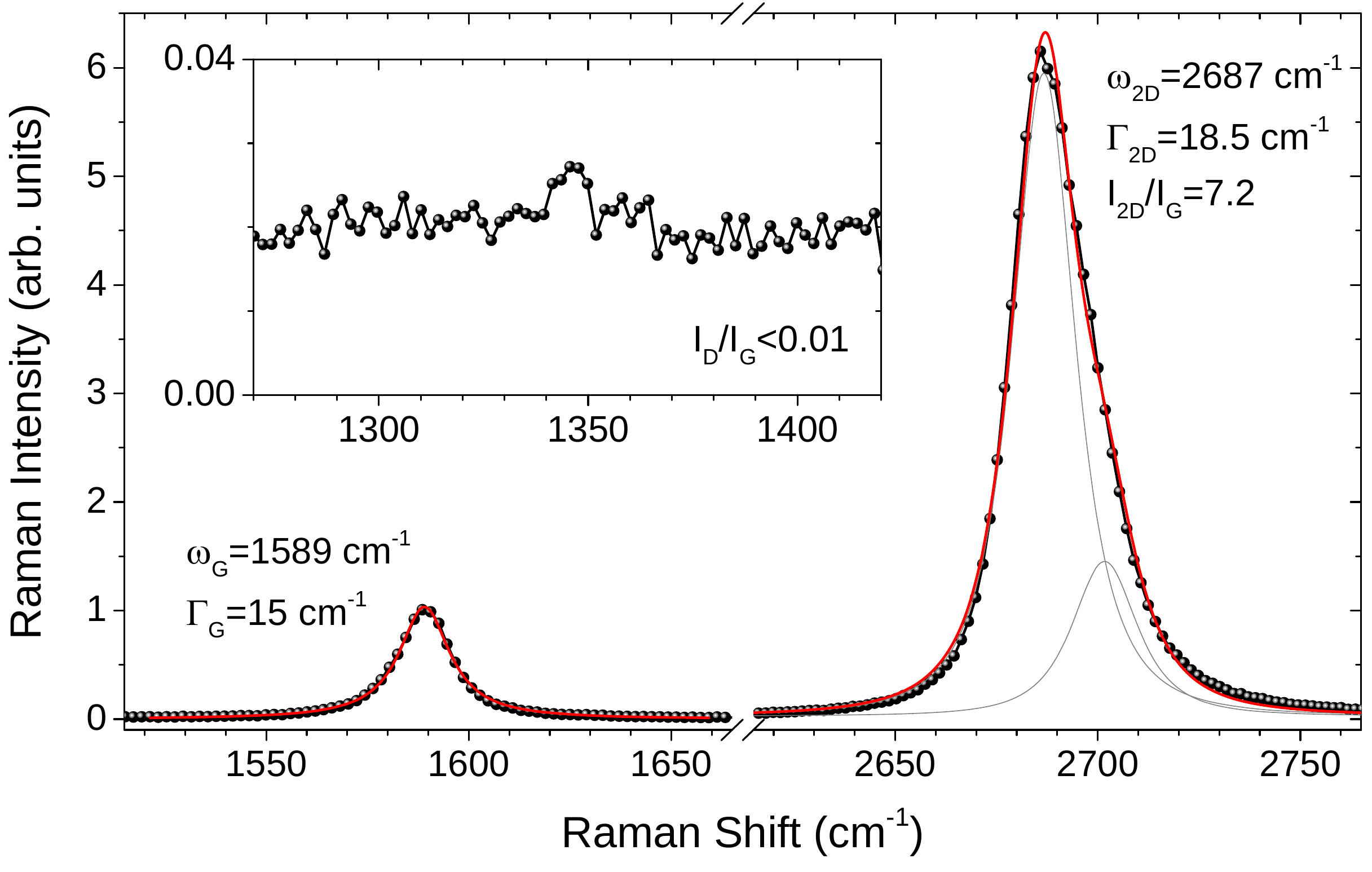}
\caption{\textbf{Micro-Raman spectrum of a freestanding graphene monolayer}. The data was recorded at $\rm B=0~\rm T$ and $T= 4 ~\rm K$, using a laser photon energy of 2.41~eV. The broad  linewidth of the G-mode feature, the large integrated intensity ratio between the 2D- and G-mode features, as well as the asymmetric lineshape of the 2D-mode feature are recognized as fingerprints of an intrinsic sample [13, 31].  The G-mode feature is fit to a Lorentzian, which peak frequency and full width at half maximum are indicated. The 2D-mode feature is fit as described in ref. 31 using a modified double-Lorentzian feature in which the two sub-features exhibit the same full width at half maximum $\Gamma_{\rm 2D}$. As shown in the inset, the very small integrated intensity ratio between the D-mode feature and the G-mode feature ($I_{\rm D}/I_{\rm G}<1\%$) indicates a very low defect contents.}

\label{SFig1}
\end{center}
\end{figure}

\begin{figure}[!ht]
\begin{center}
\includegraphics[scale=0.6]{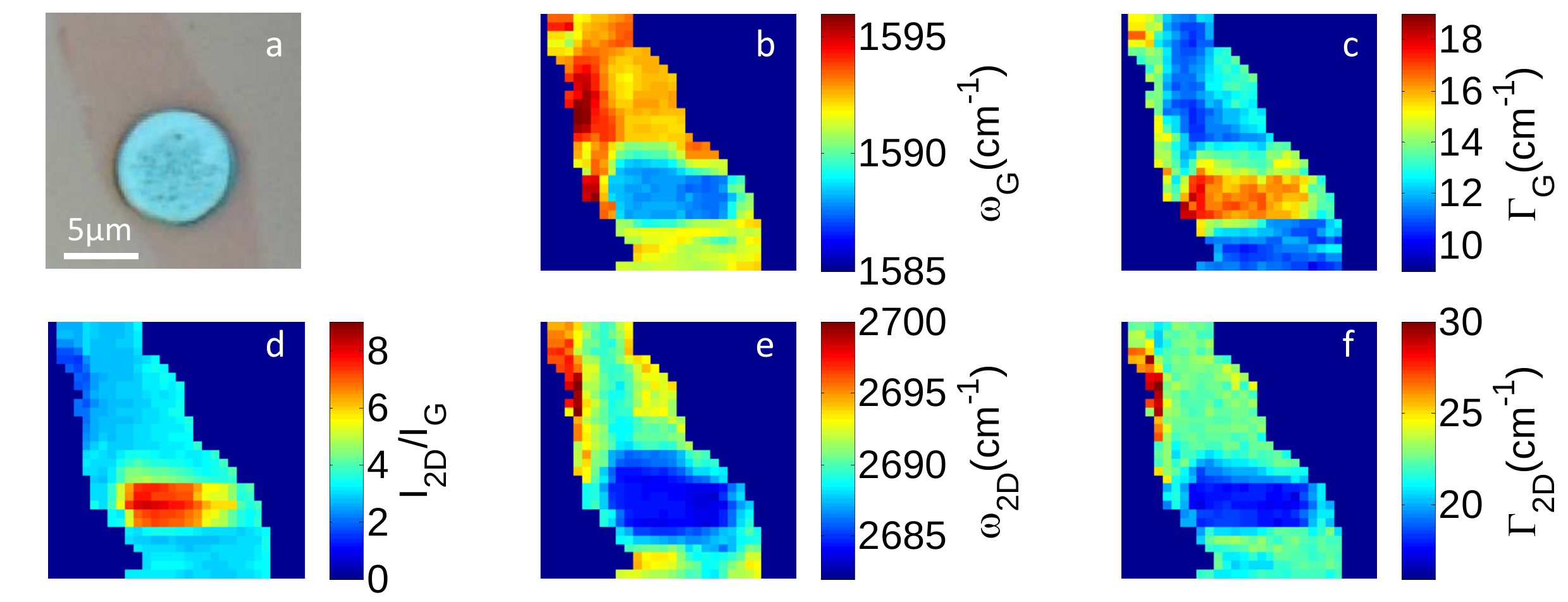}

\caption{\textbf{Spatially resolved micro-Raman study of a graphene monolayer exhibiting a freestanding region.} The data was recorded at $\rm B=0~\rm T$ and $T= 4 ~\rm K$, using a laser photon energy of 2.41~eV. a) optical image. Maps of the frequency (b) and linewidth (c) of the G-mode feature, of the integrated intensity ratio between the 2D- and G-mode features (d), of the frequency (e) and  linewidth (f) of the 2D-mode feature, as defined in Figure \ref{SFig1}. The freestanding region appears clearly on these maps and displays the Raman fingerprints of a nearly undoped sample introduced in Figure \ref{SFig1}. Over the freestanding region, the average G-mode frequency $\omega_{\rm G}^{}$ is $1587.8~ \rm cm^{-1}$, with a standard deviation of $\sim 0.5~ \rm cm^{-1}$ and the average 2D-mode frequency $\omega_{\rm 2D}^{}$ is $2684.2~ \rm cm^{-1}$, with a standard deviation of $\sim 0.5~ \rm cm^{-1}$. These values and their narrow dispersions are typical data from a very weakly doped and negligibly strained graphene monolayer, measured at low temperature.}

\label{SFig2}
\end{center}
\end{figure}

\begin{figure}[!ht]
\begin{center}
\includegraphics[scale=0.5]{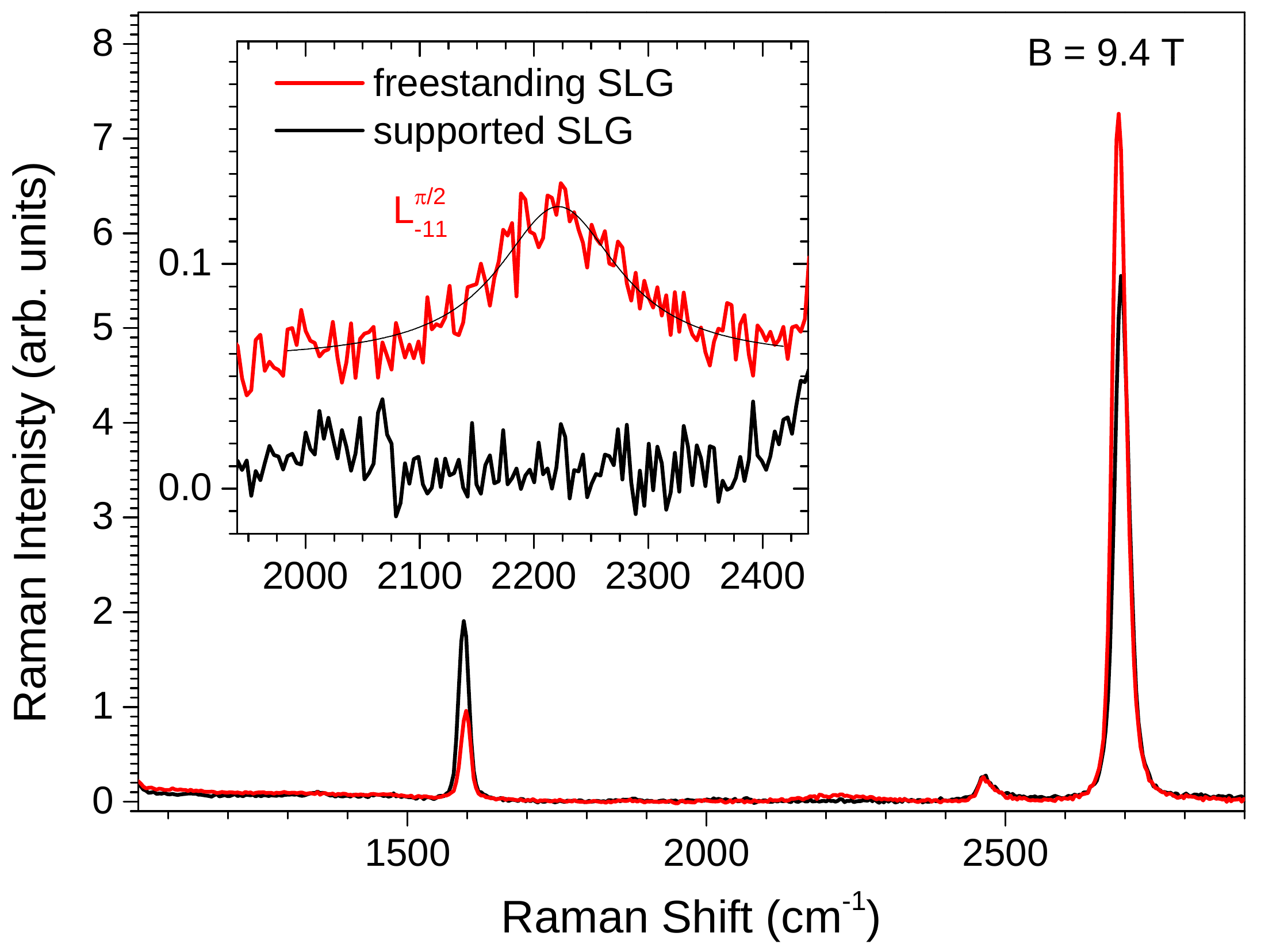}

\caption{\textbf{Mirco-magneto-Raman spectra of supported (black line) and freestanding (red line) monolayer graphene}. The data was recorded at $\rm B=9.4~\rm T$ and $T= 4 ~\rm K$, using a laser photon energy of 2.41~eV. The electronic Raman scattering feature assigned to the $L_{-1,1}^{\pi/2}$ transition appears clearly on the freestanding region but is not detectable on supported graphene. The latter feature is fit to a Lorentzian profile, with a peak frequency at $2221~\rm cm^{-1}$ and a full width at half maximum of $130~\rm cm^{-1}$.}

\label{SFig3}
\end{center}
\end{figure}

\begin{figure}[!ht]
\begin{center}
\includegraphics[scale=0.5]{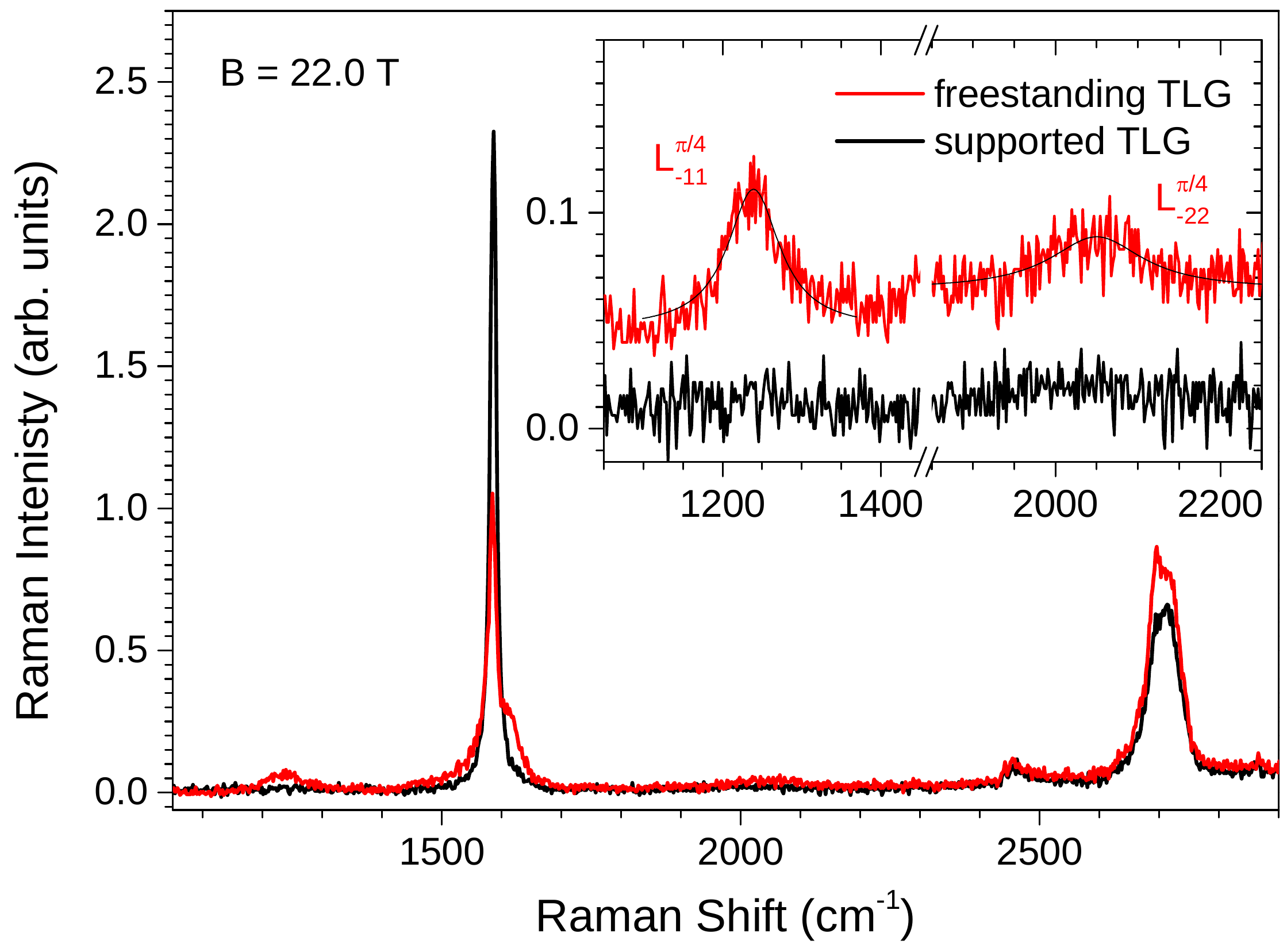}

\caption{\textbf{Mirco-magneto-Raman spectra of supported (black line) and freestanding (red line) trilayer graphene}. The data was recorded at $\rm B=22~\rm T$ and $T= 4 ~\rm K$, using a laser photon energy of 2.41~eV. The electronic Raman scattering feature assigned to the $L_{-1,1}^{\pi/4}$ and $L_{-2,2}^{\pi/4}$ transitions appears clearly on the freestanding region but are not detectable on supported graphene. The $L_{-1,1}^{\pi/4}$ ($L_{-2,2}^{\pi/4}$) feature is fit to a Lorentzian profile, with a peak frequency at $1239~\rm cm^{-1}$ ($2050~\rm cm^{-1}$) and a full width at half maximum of $80~\rm cm^{-1}$ ($137~\rm cm^{-1}$). For the sake of clarity, the $L_{-i,i}^{\pi/2}$ transitions, occurring at higher energy for $\rm B=22~\rm T$ are not shown here. These transitions can be observed in Figures \ref{Fig2} and \ref{Fig3} of the main manuscript.}

\label{SFig4}
\end{center}
\end{figure}

\begin{figure}[!ht]
\begin{center}
\includegraphics[scale=0.8]{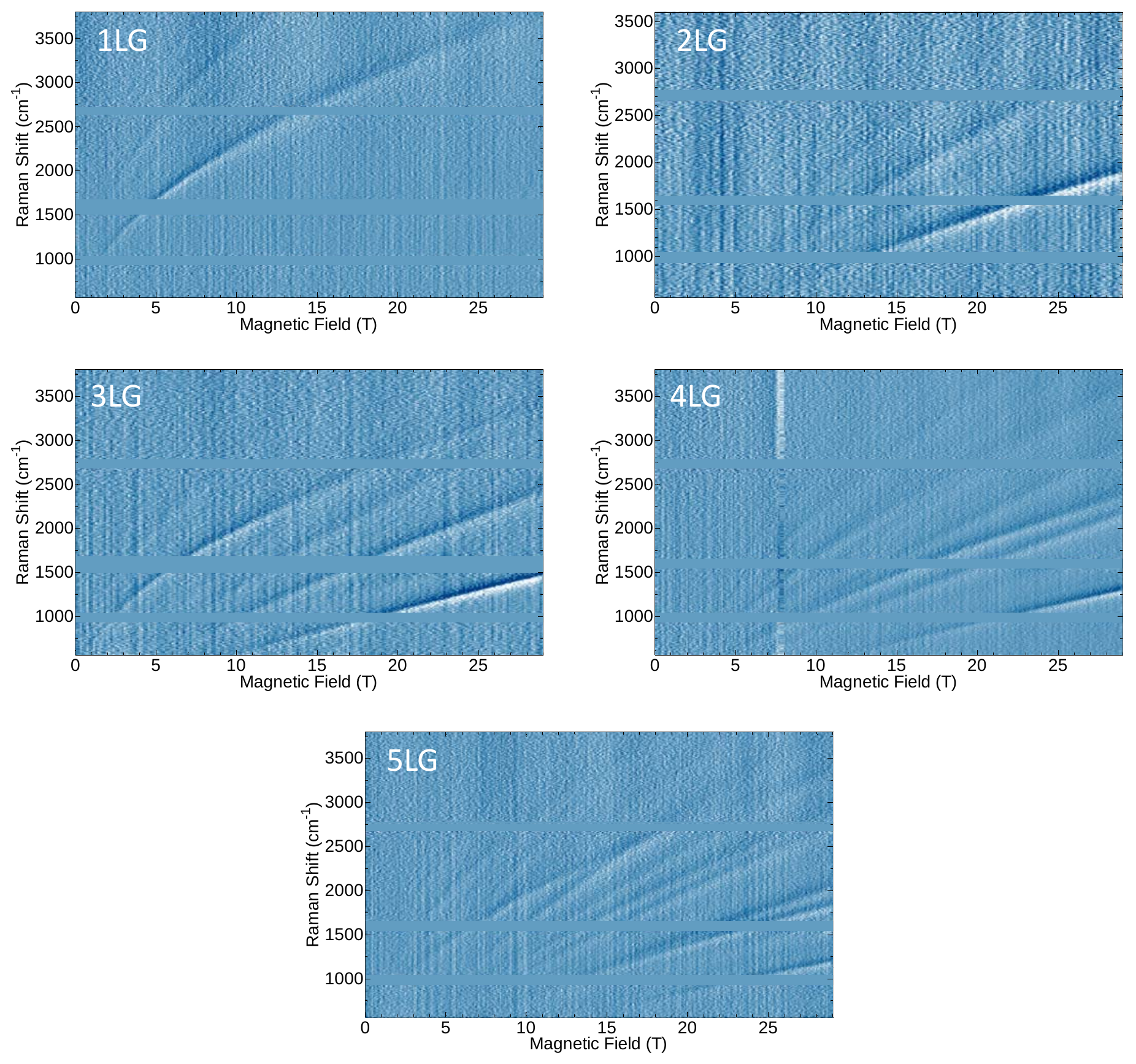}

\caption{False-color maps of the differentiated micro-Raman scattering spectra of mono- to pentalayer graphene as a function of the magnetic field. The spectrum displayed at a given field $B$ is obtained from the difference of the two Raman spectra recorded at $\rm B\pm\delta \rm B$, where $\delta\rm B$ is the magnetic field step used in each sweep. Horizontal light blue bars mask residual contributions from the G- and 2D-mode features in FLG, near $1585~\rm cm^{-1}$, and $2700~\rm cm^{-1}$, respectively, and from the underlying Si substrate, near $1000~\rm cm^{-1}$.}

\label{SFig5}
\end{center}
\end{figure}

\begin {itemize}
\item Figure \ref{SFig1} shows a typical micro-Raman spectrum of a freestanding graphene monolayer recorded at low temperature in the absence of a transverse magnetic field. 
\item Figure \ref{SFig2} shows a spatially resolved micro-Raman study of the sample in Figure \ref{SFig1}
\item Figure \ref{SFig3} shows micro-magneto-Raman scattering spectra recorded on the freestanding and supported regions of a same graphene monolayer.
\item Figure \ref{SFig4} shows micro-magneto-Raman scattering spectra recorded on the freestanding and supported regions of a same graphene triolayer.
\item Figure \ref{SFig5} shows false color maps of the differentiated micro-magneto-Raman scattering spectra of mono- to pentalayer graphene, as a function of the magnetic field.
\end {itemize}



\end{document}